\documentclass[a4paper,12pt]{article}
\usepackage{fixltx2e,fix-cm}
\usepackage{amssymb}
\usepackage{amsmath}
\usepackage{natbib}
\usepackage{graphicx}
\usepackage{subfigure}
\usepackage{makeidx}
\usepackage{multicol}

\usepackage{graphicx}
\usepackage{natbib}
\usepackage{amsmath}
\usepackage{amsfonts}
\usepackage{url}

\newcounter{ctr}

\newcounter{ctr1}

\newcounter{ctr2}

\newcounter{ctr3}

\newenvironment{theorem*}[1]{{\bf Theorem #1} \begin{itshape}}{\end{itshape}}

\newenvironment{corollary*}[1]{{\bf Corollary #1} \begin{itshape}}{\end{itshape}}

\newenvironment{proposition*}[1]{{\bf Proposition #1} \begin{itshape}}{\end{itshape}}

\newcommand{\ud}{\, {\rm d} \kern-.015em }

\newcommand{\modulus}[1]{\left| \kern.05em #1 \kern.05em \right|}
\newcommand{\norm}[1]{\left\| \kern.05em #1 \kern.05em \right\|}
\newcommand{\inner}[1]{\left\langle \kern.05em #1 \kern.05em \right\rangle }

\newcommand{\pick}[2]{\renewcommand{\arraystretch}{0.6}
\left( \kern-.4em \begin{array}{c} #1 \\ #2 \end{array} \kern-.4em \right) }

\usepackage{authblk}

\makeindex

\def\bs{\pmb{s}}
\def\piABC{\pi_{ABC}}
\def\bx{\pmb{x}}
\def\bb{\pmb{b}}
\def\bZ{\pmb{Z}}
\def\bV{\pmb{V}}
\def\bt{\pmb{t}}
\def\balpha{\pmb{\alpha}}
\def\bS{\pmb{S}_{n,\btheta}}
\def\tbS{\tilde{\pmb{S}}_{n,\btheta}}
\def\btheta{\pmb{\theta}}
\def\sobs{\bs_{obs}}
\def\sobss{s_{obs}}
\def\tsobs{\tilde{\bs}_{obs}}

\title{Asymptotics of ABC}
\date{}
\author[1,$\dag$]{Paul Fearnhead}
\affil[1]{Department of Mathematics and Statistics, Lancaster University}
\affil[$\dag$]{Correspondence: p.fearnhead@lancaster.ac.uk}

\begin{document}
\maketitle 

%
\begin{center}
 {\bf Abstract}
 \end{center}
This document is due to appear as a chapter of the forthcoming Handbook of Approximate Bayesian Computation (ABC) edited by S. Sisson, Y. Fan, and M. Beaumont. 

We present an informal review of recent work on the asymptotics of Approximate Bayesian
Computation (ABC). In particular we focus on how does the ABC posterior, or point estimates obtained by ABC, behave in the limit as we have more data? 
The results we review show that ABC can perform well in terms of point estimation, but standard implementations will over-estimate the uncertainty about the parameters. If we use
the regression correction of Beaumont et al. then ABC can also accurately quantify this uncertainty. The theoretical results also have practical implications for how to implement
ABC. 
 
\section{Introduction}\label{intro}

This chapter aims to give an overview of recent work on the asymptotics of Approximate Bayesian
Computation (ABC). By asymptotics here we mean how does the ABC posterior, or point estimates obtained by ABC, behave in the limit as we have more data? The chapter summarises results from three papers,
\cite{li2015}, \cite{Frazier:2016} and \cite{li2016}. The presentation in this chapter is deliberately informal, with the hope of conveying both the intuition behind the theoretical results from these papers 
and the practical consequences of this theory. As such we will not present all the technical conditions for the results we give: 
the interested reader should consult the relevant papers for these, and the results we state should be interpreted as holding under appropriate regularity conditions.

We will focus on ABC for a $p$-dimensional parameter, $\btheta$, from a prior $p(\btheta)$ (we use the common convention of denoting vectors in bold, and we will assume these are column vectors). 
We assume we have data of size $n$ that is summarised through a $d$-dimensional summary statistic. The asymptotic results we review consider the limit $n\rightarrow\infty$, but assume that the summary
statistic is of fixed dimension. Furthermore all results assume that the dimension of the summary statistic is at least as large as the dimension of the parameters, $d\geq p$ -- this is implicit in the identifiability conditions that we will introduce later. Examples of such a setting are where the summaries are sample means of functions of individual data points, quantiles of the data, or, for time-series data, are empirical auto-correlations of the data. It also includes summaries based on fixed-dimensional auxillary models \cite[]{drovandi2015bayesian} or on composite likelihood score functions \cite[]{ruli2016approximate}.

To distinguish the summary statistic for the observed data from the summary statistic of data simulated within ABC, we will denote the former by $\sobs$, and the latter by $\bs$. Our model for the data will define a probability model for the summary. We assume that this in turn specifies a probability density
function, or likelihood, for the summary, $f_n(\bs;\btheta)$, which depends on the parameter. 
In some situations we will want to refer to the random variable for the summary statistic, and this will be $\bS$. As is standard with ABC, we assume that we can simulate from the model but cannot calculate
$f_n(\bs;\btheta)$.

The most basic ABC algorithm is a rejection sampler \cite[]{Pritchard:1999} \index{ABC Rejection Sampler}, which iterates the following
three steps:
\begin{itemize}
\item[(RS1)] Simulate a parameter from the prior: $\theta_i\sim p(\btheta)$.
\item[(RS2)] Simulate a summary statistic from the model given $\btheta_i$: $\bs_i\sim f_n(\bs|\btheta_i)$.
\item[(RS3)] Accept $\btheta_i$ if $\lVert \sobs-\bs_i \rVert<\epsilon$.
\end{itemize}
Here $\lVert \sobs-\bs_i \rVert$ is a suitably chosen distance between the observed and simulated summary statistics, and $\epsilon$ is a suitably chosen bandwidth. In the following we will assume that $||\bx||$ is either Euclidean distance, $||\bx||^2=\bx^T\bx$, or a Mahalanobis distance, $||\bx||^2=\bx^T \Gamma \bx$ for some chosen positive-definite $d\times d$ matrix $\Gamma$. 

If we define a (uniform) kernel function, $K(\bx)$, to be 1 if $\lVert \bx \rVert<1$
and 0 otherwise, then this rejection sampler is drawing from the following distribution
\[
\piABC(\btheta)\propto p(\btheta) \int f_n(\bs|\btheta) K\left( \frac{\sobs-\bs}{\epsilon} \right) \mbox{d}\bs.
\]
We call this the ABC posterior \index{ABC Posterior}. If we are interested in estimating a function of the parameter $h(\btheta)$ we can use the ABC posterior mean
\[
h_{ABC}=\int h(\btheta)\piABC(\btheta)\mbox{d}\btheta.
\]
In practice we cannot calculate this posterior mean \index{ABC Posterior Mean} analytically, but would have to estimate it based on the sample mean of $h(\btheta_i)$ for parameter values $\btheta_i$ simulated using the above rejection sampler. 

In this chapter we review results on the behaviour of the ABC posterior, the ABC posterior mean, and Monte Carlo estimates of this mean as $n\rightarrow\infty$. In particular we consider whether the ABC posterior concentrates around the true parameter value in Section \ref{S:Concentration}. We then consider the limiting form of the ABC posterior and the frequentist asymptotic distribution of the ABC posterior mean in Section \ref{S:Posterior}. For the latter two results we compare these asymptotic
distributions with those of the true posterior given the summary -- which is the best we can hope for once we have chosen our summary statistics. 

The results in these two sections ignore any Monte Carlo error. The impact of Monte Carlo error on the asymptotic variance of our ABC posterior mean estimate is the focus of Section \ref{S:MonteCarlo}. This impact depends on the choice of algorithm we use
to sample from the ABC posterior (whereas the choice of algorithm has no effect on the actual ABC posterior or posterior mean that are analysed in the earlier sections). The rejection sampling algorithm above is inefficient in the
limit as $n\rightarrow\infty$ and thus we consider more efficient importance sampling and MCMC generalisations in this section.
 
We then review results that show how post-processing the output of ABC can lead to substantially stronger asymptotic results. The chapter then finishes with a discussion that aims to draw out the
key practical insights from the theory.

Before we review these results, it is worth mentioning that we can generalise the definition of the ABC posterior, and the associate posterior mean, given above.
Namely we can use a more general form of kernel than the uniform kernel. Most of the results we review apply if we replace the uniform kernel by
a different kernel, $K(\bx)$, that is monotonically decreasing in $\lVert \bx \rVert$. Furthermore the specific form of the kernel has little affect on
the asymptotic results -- what matters most is how we choose the bandwidth and, in some cases, the choice of distance. The fact that most of the theoretical results do not depend on the choice of kernel means that, 
for concreteness, we will primarily assume a uniform kernel in our presentation below. The exceptions being in Section \ref{S:Posterior} where it is easier to get an intuition for 
the results if we use a Gaussian kernel. By focussing on these two choices we do not mean to suggest that they are necessarily better than other choices, it is just that they simplify the exposition. We will return to the choice of kernel in the Discussion.

\section{Posterior Concentration} \label{S:Concentration} \index{Posterior Concentration}

The results we present in this section are from \cite{Frazier:2016} \cite[though see also][]{martin2014approx}, and consider the question of whether the ABC posterior will place increasing probability mass around the true parameter value as $n\rightarrow\infty$. It is the most basic convergence result we would wish for, requires weaker conditions than results we give in Section \ref{S:Posterior}, and is thus easier to
apply  to other ABC settings \cite[see for example][]{marin2014relevant,bernton2017inference}.

We will denote the true parameter value by $\btheta_0$. If we define
\[
\mbox{Pr}_{ABC}( \lVert \btheta-\btheta_0 \rVert<\delta)=\int_{\btheta : \lVert \btheta-\btheta_0 \rVert<\delta} \piABC(\btheta)\mbox{d}\btheta,
\]
the ABC posterior probability that $\btheta$ is within some distance $\delta$ of the true parameter value, then for posterior concentration we
want that for any $\delta>0$\index{Posterior Concentration}
\[
\mbox{Pr}_{ABC}( \lVert \btheta-\btheta_0 \rVert<\delta)\rightarrow 1
\]
as $n\rightarrow \infty$. That is, for any strictly positive choice of distance, $\delta$, regardless of how small it is, as $n\rightarrow\infty$ we need the ABC posterior to place all its probability on the event that $\btheta$ is within $\delta$ of the true parameter value.

To obtain posterior concentration for ABC we will need to let the bandwidth depend on $n$, and henceforth we denote the bandwidth by $\epsilon_n$. 

\subsection{ABC Posterior Concentration} 

The posterior concentration result of \cite{Frazier:2016} is based upon assuming a law of large numbers for the summary statistics. Specifically we need the existence of a binding function, $\bb(\btheta)$, such that for any $\btheta$
\[
\bS \rightarrow \bb(\btheta)
\]
\index{Binding Function}
in probability as $n\rightarrow\infty$. If this holds, and the binding function satisfies an identifiability \index{Identifiability} condition: that $\bb(\btheta)=\bb(\btheta_0)$ implies $\btheta=\btheta_0$, then we have posterior concentration providing the bandwidth tends to zero, $\epsilon_n\rightarrow 0$.

\begin{figure}
\centering
\includegraphics[scale=0.45]{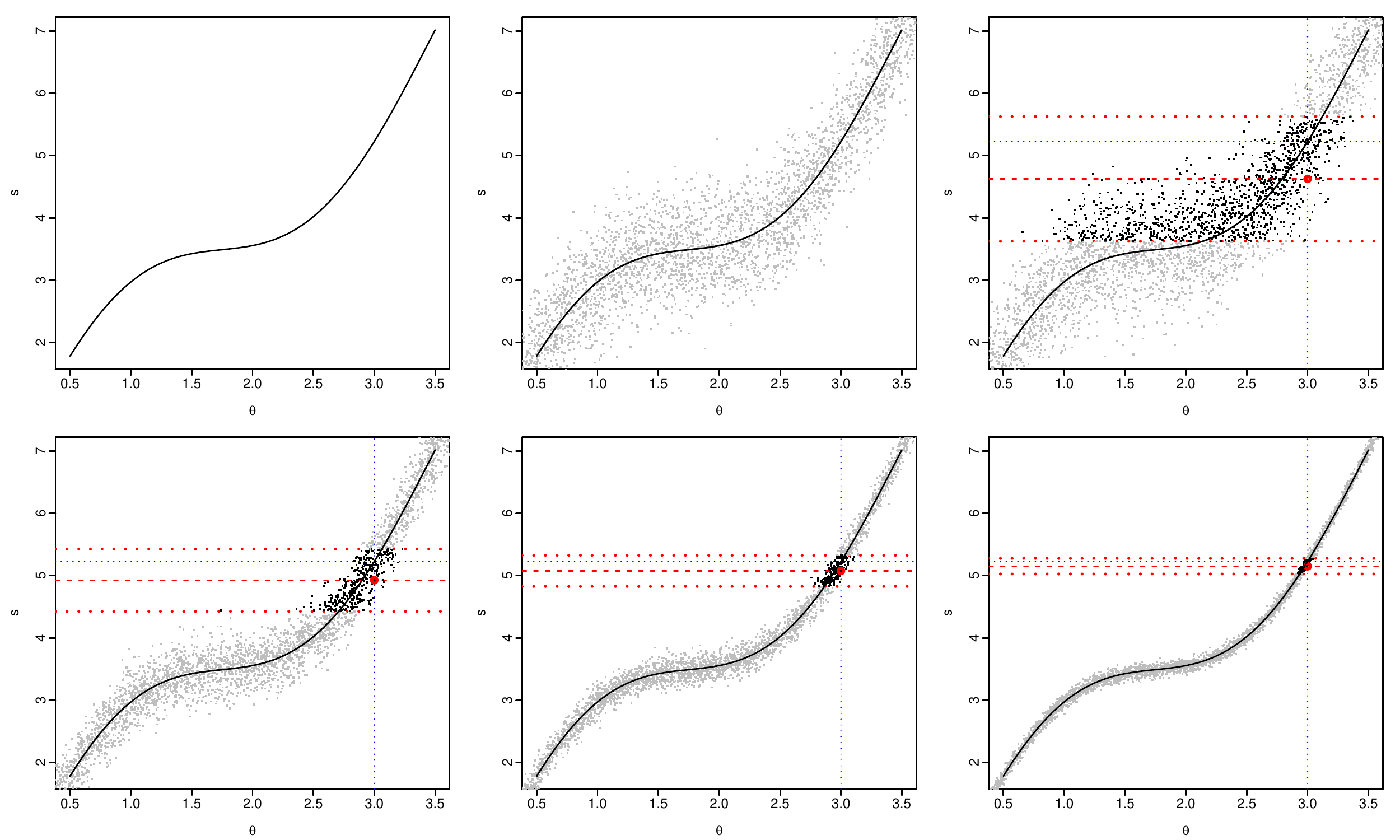}
\caption{
Example binding function, $b(\theta)$ (top-left plot). Pairs of parameter and summary statistic values proposed by
a rejection sampler (top-middle). Output of rejection sampler (top-right): $\theta_0$ and $b(\theta_0)$ (blue dotted vertical and horizontal lines respectively); 
$\sobss$ (bold red circle, and red dashed horizontal line) and acceptance region for proposed summaries (bold red dashed horizonal lines); pairs of parameter and summary statistic values accepted (bold)
and rejected (grey) by the rejection sampler. Bottom-row plots are the same as top-right plot but for increasing $n$ and decreasing $\epsilon_n$. Here, and for all plots, our results are for a 
simple scenario where data is IID Gaussian with a mean that is a function of the parameter, and the summary statistic is the sample mean. 
(In this case the binding function is, by definition, equal to the mean function.)
\label{Fig:concentration1}
}
\end{figure}

To gain some insight into this result and the assumptions behind it, we present an example. To be able to visuallise what is happening we will assume that the parameter and summary statistic are both 1-dimensional. Figure \ref{Fig:concentration1}  shows an example binding function, a value of $\theta_0$ and $\sobss$, and output from the ABC rejection sampler. 

As $n$ increases we can see the plotted points, that show proposed parameter and summary statistic values, converge
towards the line that shows the binding function. This stems from our assumption of a law of large numbers for the summaries,
so that for each $\theta$ value the summaries should tend to $b(\btheta)$ as $n$ increases. 

We also have that the observed summary statistic, $\sobss$, converges towards
$b(\theta_0)$. Furthermore we are decreasing the bandwidth as we increase $n$, which corresponds to narrower acceptance regions for the summaries, which  means that the accepted summary statistics converge towards 
$b(\theta_0)$. Asymptotically, only parameter values close to $\theta_0$, which have values $b(\theta)$ which are
close to $b(\theta_0)$, will simulate summaries close to $b(\theta_0)$. Hence the only accepted parameter values will be 
close to, and asymptotically will concentrate on, $\theta_0$. This can be seen in practice from the plots in the bottom
row of Figure \ref{Fig:concentration1}.

The identifiability \index{Identifiability} condition on the binding function is used to ensure that concentration of accepted summaries around
$\bb(\btheta_0)$ results in ABC posterior concentration around $\btheta_0$. What happens when this identifiability condition does not hold is discussed in Section \ref{S:bind}.
\index{Binding Function}
\index{Posterior Concentration}

\subsection{Rate of Concentration} \label{S:Rate}

We can obtain stronger results by looking at the rate at which concentration occurs. Informally we can think of this  as the supremum of rates, $\lambda_n\rightarrow0$, such that \index{Posterior Concentration Rate}
\[
\mbox{Pr}_{ABC}( \lVert \btheta-\btheta_0 \rVert <\lambda_n)\rightarrow 1
\]
as $n\rightarrow\infty$.  For parametric Bayesian inference with independent and identically distributed data this rate would be $1/\sqrt{n}$.

Assuming the binding function is continuous at $\btheta_0$, then the rate of concentration will be determined by the rate at which accepted summaries concentrate on $\bb(\btheta_0)$. As described above, this depends on the variability (or `noise') 
of the simulated summaries around the binding function and on the bandwidth, $\epsilon_n$. The rate of concentration will be the slower of the rate at which the noise in the summary statistics and the rate at
which $\epsilon_n$ tend to 0. 

\begin{figure}
\centering
\includegraphics[scale=0.45]{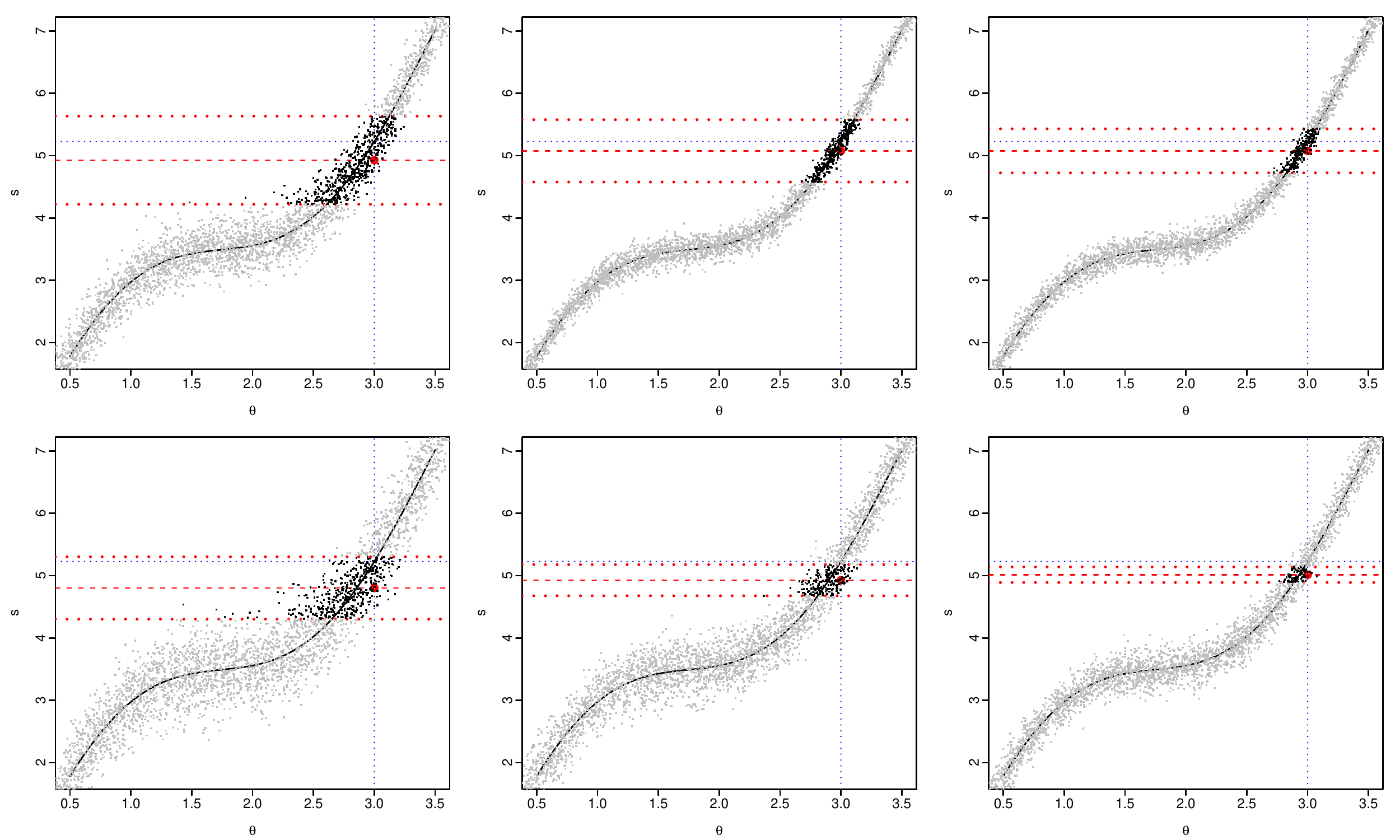}
\caption{
Example of ABC concentration for differing rates of the noise in the summary statistics and rates of $\epsilon_n$. Plots are as in Figure \ref{Fig:concentration1}. Top-row: noise in summary statistics halving, or equivalently sample size increasing by a factor of 4,  while  $\epsilon_n$ decreasing by $1/\sqrt{2}$ as we move from left to right. Bottom-row: noise in summary statistics decreasing by  $1/\sqrt{2}$, or equivalently sample size doubling,  while  $\epsilon_n$ halving as we move from left to right. 
\label{Fig:Rate}
}\index{Posterior Concentration Rate}
\end{figure}

We can see this from the example in Figure \ref{Fig:Rate}, where we show output from the ABC rejection sampler for different values of $n$, but with $\epsilon_n$ tending to 0 at either a faster or slower rate than that of the noise in the summaries. For each regime the rate of concentration of both the accepted summaries and of the accepted parameter values is determined by the slower of the two rates.

\subsection{Effect of Binding Function} \label{S:bind} \index{Binding Function}

\begin{figure}
\centering
\includegraphics[scale=0.45]{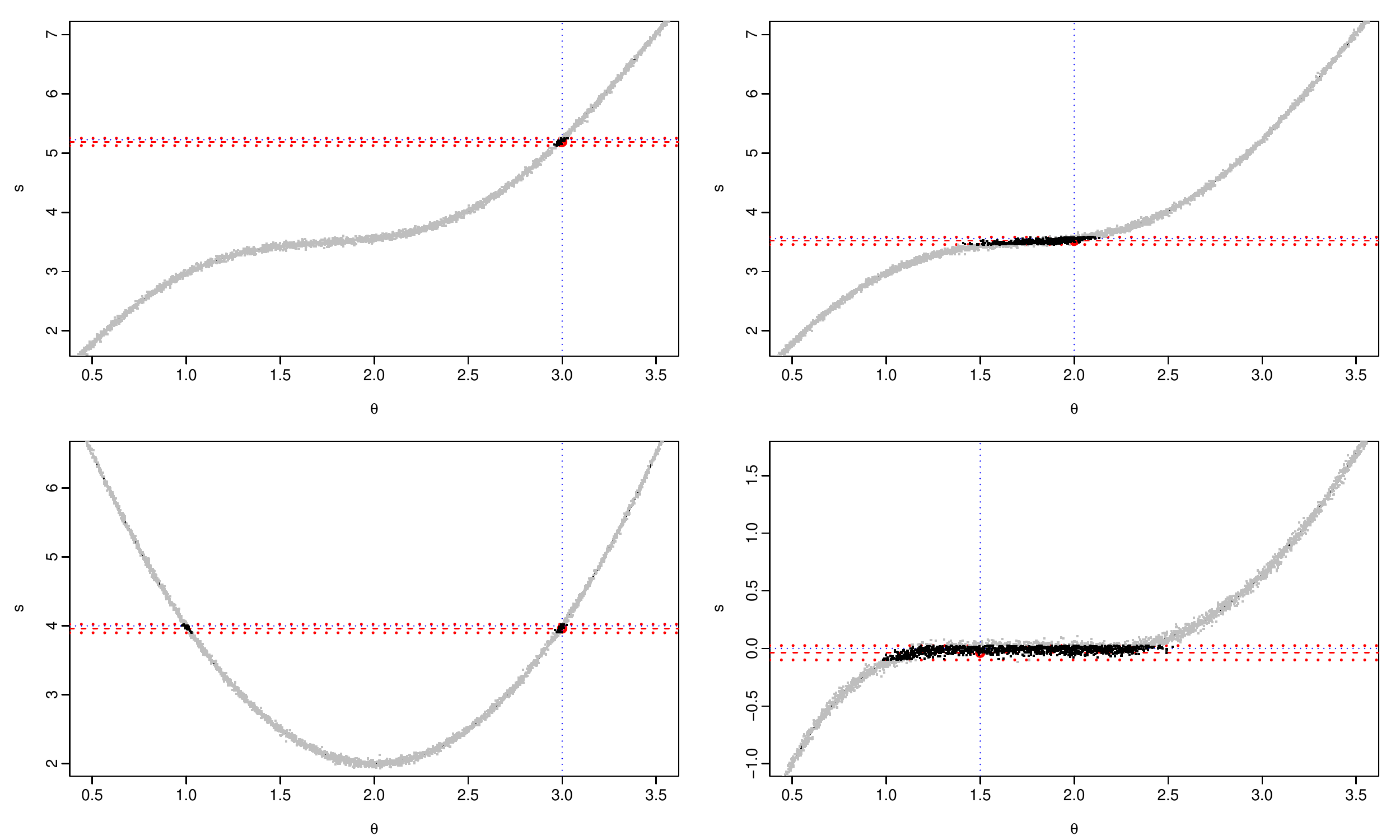}
\caption{
Example of the effect of the shape of binding function on the ABC posterior (plots are as in Figure \ref{Fig:concentration1}).
 Top row: gradient of binding function at
$b(\theta_0)$ affects the ABC posterior variance, with larger gradient (left-hand plot) resulting in lower ABC posterior variance
than smaller gradient (right-hand plot). Bottom row: effect of non-identifiability on ABC posterior.
\label{Fig:Bind}
}
\end{figure}

The shape of the binding function for values of $\btheta$ for which $\bb(\btheta)$ is close to $\bb(\btheta_0)$ affects the ABC posterior as it affects the range of $\btheta$ values that will have a
reasonable chance of producing summary statistic values that would be accepted by the ABC rejection sampler. 

If the identifiability condition \index{Identifiability} holds and the binding function is differentiable at $\btheta_0$ then the value of this gradient will directly impact the ABC posterior variance. This is shown in the top row of Figure \ref{Fig:Bind}. If this gradient is large (top-left plot) then even quite large differences in summary statistics would correspond to small differences in the parameter, and hence a small ABC posterior variance. By comparison if the gradient is small (top-right plot) then large differences in parameters may mean only small differences in summary statistics. In this case we expect a much larger ABC posterior variance for the same width of the region in which the summary statistics are accepted.

The bottom row of Figure \ref{Fig:Bind} shows what can happen if the identifiability condition does not hold. The bottom-left plot gives an example where there are two distinct parameter values for which the binding function is equal to $b(\theta_0)$. In this case we have a bi-modal ABC posterior that concentrates on these two values. The bottom-right plot shows an example where there is a range of parameter values whose binding function value is equal to $b(\theta_0)$, and in this case the ABC 
posterior will concentrate on this range of parameter values. 

It can be difficult in practice to know whether the identifiability condition \index{Identifiability} holds. In large data settings, observing a multi-modal posterior as in the bottom-left plot of Figure \ref{Fig:Bind} would suggest that it does not hold. In such cases it may be possible
to obtain identifiability by adding extra summaries. The wish to ensure identifiability is one reason for choosing a higher dimensional summary than parameter. However this does not come without potential cost, as we show in Section \ref{S:Posterior}.

\subsection{Model Error} \label{S:ModelError} \index{Model Error} \index{Binding Function}

\begin{figure}
\centering
\includegraphics[scale=0.45]{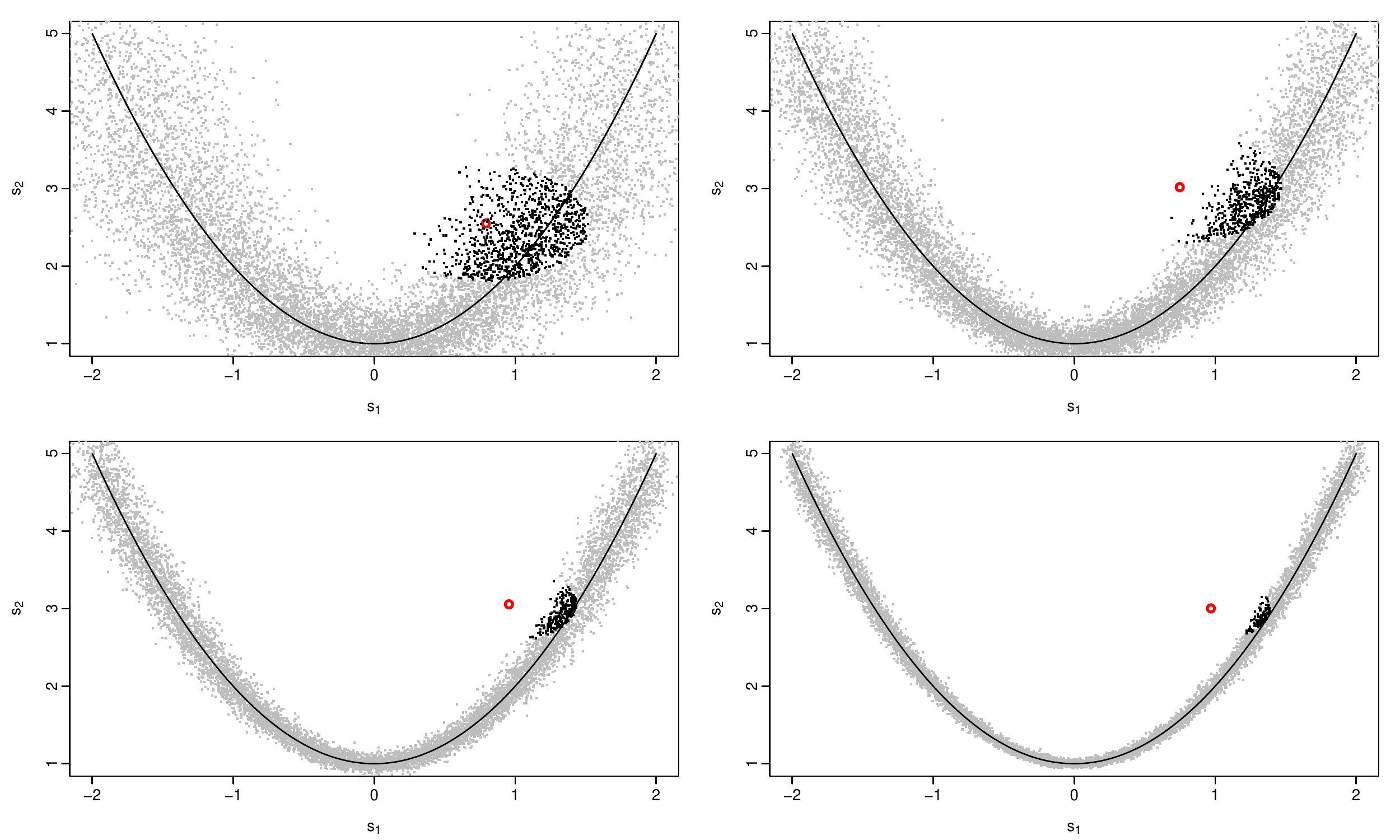}
\caption{
Example of the effect of model error in ABC for the Gaussian model with incorrect variance described in the text. The plots, from left to right and top to bottom, correspond to increasing sample size. Each plot shows the 2-dimensional binding function as we vary $\theta$ (line); the observed summary statistic (red circle) and accepted (black dots) and rejected (grey dots) summary statistic values. (For this model the parameter value used to simulate the summary statistics will be close to the first summary
statistic, $s_1$.) \index{Model Error}
\label{Fig:ModelError}
}
\end{figure}

One of the implicit assumptions behind the result on posterior concentration is that our model is correct. This manifests itself within the assumption that as we get more data the observed summary statistic will converge to the value $\bb(\btheta_0)$. If the model we assume in ABC is incorrect then this may not be the case \cite[see][for a fuller discussion of  the impact of model error]{Frazier:2017}.  There are then two possibilities, the first is that the observed summary statistic will converge to a value $\bb(\tilde{\btheta})$ for some parameter value $\tilde{\btheta}\neq \btheta_0$. 
 In this case, by the arguments above, we can still expect posterior concentration but to $\tilde{\btheta}$ and not $\btheta_0$.

The other possibility is that the observed summary statistic converges to a value that is not equal to $\bb(\btheta)$ for any $\btheta$. This is most likely to occur when the dimension of the summary statistic is greater than the dimension of the parameter. To give some insight into this scenario, we give in an example in Figure \ref{Fig:ModelError}, where we have independent identically distributed data from a Gaussian distribution with mean $\theta$ and variance $\theta^2+2$, but our
model assumes the mean and variance are $\theta$ and $\theta^2+1$ respectively. This corresponds to a wrong assumption about the variance. We then apply ABC with summary statistics that are the sample mean and variance.

As shown in the figure, we still can get posterior concentration in this setting. If we denote the limiting value of the binding function for the true model as $b_0$, then the posterior concentrates on parameter value, or values, whose binding function value is closest, according to the distance we use for deciding whether to accept simulated summaries, to $b_0$. 

In this second scenario it may be possible to detect the model error by monitoring the closeness of the accepted summaries to the observed summaries. If the model is correct, then the distance between accepted and observed summaries tends to 0 with increasing $n$. Whereas in this second model error scenario, these distances will tend towards some non-zero constant.

\section{ABC Posterior and Posterior Mean} \label{S:Posterior}

We now consider stronger asymptotic results for ABC. To obtain these results we need extra assumptions in addition to those required for posterior concentration \cite[see][for full details]{Frazier:2016,li2015}. The most important of these is that the summary statistics obey a central limit theorem
\[
\sqrt{n}\left\{\bS-\bb(\btheta)\right\}
\rightarrow
\mbox{N}\left\{
0,A(\btheta)
\right\},
\]
for some $d\times d$ positive definite matrix $A(\btheta)$. In the above central limit theorem we have assumed a $1/\sqrt{n}$ rate of convergence, but it is trivial to generalise this \cite[]{li2015}.

\subsection{ABC Posterior} \label{S:Posterior_sub} \index{ABC Posterior}

Under this central limit assumption we first consider convergence of the ABC posterior. Formal results can be found in \cite{Frazier:2016} \cite[but see also][]{li2016}. Here we give an informal presentation of these results. 

To gain intuition about the limiting form of the ABC posterior, we can use the fact from the previous section that there is posterior concentration around $\btheta_0$. Thus asymptotically we need only consider the behaviour of the model for $\btheta$ close to $\btheta_0$. Also asymptotically the noise in the summaries is Gaussian. So if we make a linear approximation to
$\bb(\btheta)$ for $\btheta$ close to $\btheta_0$, our model will be well approximated by
\[
\bS=\bb(\btheta_0)+D_0 (\btheta-\btheta_0)+\frac{1}{\sqrt{n}}\bZ,
\]
where $D_0$ is the $d\times p$ matrix of first derivatives of $\bb(\btheta)$ with respect to $\btheta$, with these 
derivatives evaluated at $\btheta_0$; and $\bZ$ is a $d$-dimensional Gaussian random variable with covariance matrix
$A(\btheta_0)$. Furthermore, for $\btheta$ close to $\btheta_0$ the prior will be well approximated by a uniform prior.
For the following we assume that $D_0$ is of rank $p$.

\cite{wilkinson2013approximate} shows that the effect of the approximation in ABC, whereby we accept simulated summaries which are similar, but not identical, to the
observed summary, is equivalent to performing exact Bayesian inference under a different model. This different model has additional additive noise, where the distribution of the noise is given by the kernel,  $K(\cdot)$, we use in ABC. So if $\bV$ is 
a $d$-dimensional random variable with density $K(\cdot)$, independent of $\bZ$, then our ABC posterior \index{ABC Posterior Asymptotics} will behave like the true posterior for the 
model
\begin{equation} \label{eq:approx}
\bS=\bb(\btheta_0)+D_0 (\btheta-\btheta_0)+\frac{1}{\sqrt{n}}\bZ+\epsilon_n\bV.
\end{equation}

From Section \ref{S:Rate}, we know that the rate of concentration is the slower of the rate of
the noise in the summaries, $1/\sqrt{n}$ under our central limit theorem, and the bandwidth $\epsilon_n$. This means that we get different limiting results depending on whether $\epsilon_n=O(1/\sqrt{n})$ or not. This can be seen from (\ref{eq:approx}), as
whether $\epsilon_n=O(1/\sqrt{n})$ or not will affect whether the $\epsilon_n\bV$ noise term dominates or not.

If $\sqrt{n}\epsilon_n\rightarrow \infty$, so $\epsilon_n$ is the slower rate, then to get convergence of the ABC posterior we
need to consider the re-scaled variable $\bt= (\btheta-\btheta_0)/\epsilon_n$. If we further define $\tbS=\{\bS-\bb(\btheta_0)\}/\epsilon_n$ then we can re-write (\ref{eq:approx}) as
\[
\tbS=D_0 \bt + \bV + \frac{1}{\epsilon_n\sqrt{n}}\bZ\rightarrow D_0 \bt + \bV.
\]
Thus the limiting form of the ABC posterior is equivalent to the true posterior for this model, given observation $\tsobs=\{\sobs-\bb(\btheta_0)\}/\epsilon_n$, with a uniform prior for $\bt$. The shape of this posterior will be determined by the ABC kernel.
If we use the standard uniform kernel, then the ABC posterior \index{ABC Posterior} will asymptotically be uniform. By converting from $\bt$ to $\btheta$ we see that the asymptotic variance for
$\btheta$ is $O(1/\epsilon_n^2)$ in this case.

The other case is that $\sqrt{n}\epsilon_n\rightarrow c$ for some positive, finite constant $c$. In this case we consider
the re-scaled variable $\bt= \sqrt{n}(\btheta-\btheta_0)$, and re-scaled observation $\tbS=\sqrt{n}\{\bS-\bb(\btheta_0)\}$. The ABC posterior \index{ABC Posterior Asymptotics} will asymptotically be equivalent to the true posterior 
for $\bt$ under a uniform prior, for a model
\[
\tbS=D_0 \bt + \bZ + {\epsilon_n\sqrt{n}}\bV\rightarrow D_0 \bt + \bZ+c\bV,
\]
and given an observation $\tsobs=\sqrt{n}\{\sobs-\bb(\btheta_0)\}$. 

We make three observations from this. First if $\epsilon_n=o(1/\sqrt{n})$, so $c=0$, then using standard results for the posterior distribution of a linear model, the ABC posterior for $\bt$ will converge to a Gaussian with mean
\begin{equation} \label{eq:Mean1}
\left\{D_0^TA(\theta_0)^{-1}D_0\right)\}^{-1}D_0^TA(\theta_0)^{-1}\tsobs,
\end{equation}
and variance $I^{-1}$ where $I=D_0^TA(\btheta_0)^{-1}D_0$. This is the same limiting form as the true posterior given the summaries. The matrix $I$ can be viewed as an information matrix, and note that this is larger if the derivatives of the binding
function, $D_0$, are larger; in line with the intuition we presented in Section \ref{S:bind}.

Second if $c\neq0$, the ABC posterior will have a larger variance than the posterior given summaries. This inflation of the ABC
posterior variance will increase as $c$ increases. In general it is hard to say the form of the posterior, as it will depend on the
distribution of noise in our limiting model, $\bZ+c\bV$, which is a convolution of the limiting Gaussian noise of the summaries and a random variable drawn from the ABC kernel.

Our final observation is that we can get some insight into the behaviour of the ABC posterior when $c\neq0$ if we assume a Gaussian kernel, as again the limiting ABC posterior \index{ABC Posterior Asymptotics} will be the true posterior for a linear a model with Gaussian noise. If the Gaussian kernel has variance $\Sigma$, which corresponds to measuring distances between summary statistics using the scaled distance $\lVert \bx \rVert=\bx^T \Sigma^{-1} \bx$, then the ABC posterior for $\bt$ will converge to a Gaussian with
mean
\begin{equation} \label{eq:Mean2}
\left\{D_0^T(A(\theta_0)+c^{2}\Sigma)^{-1}D_0\right\}^{-1}D_0^T\{A(\theta_0)+c^2\Sigma\}^{-1}\tsobs
\end{equation}
and variance, $\tilde{I}^{-1}$, where
\[
\tilde{I}=D_0^T\{A(\theta_0)+c^{2}\Sigma\}^{-1}D_0.
\]

\subsection{ABC Posterior Mean} \index{ABC Posterior Mean}

We now consider the asymptotic distribution of the ABC posterior mean \index{ABC Posterior Mean Asymptotics}. By this we mean the frequentist distribution, whereby we view the posterior mean as a function of the data, and look at the distribution of this under repeated sampling of the data.
 Formal results appear in \cite{li2015}, but we will give informal results, building on the results we gave for the ABC posterior. 
 We will focus on the case where $\epsilon_n=O(1/\sqrt{n})$, but note that results hold for the situation where $\epsilon_n$ decays more slowly; in fact \cite{li2015} show that if $\epsilon_n=o(n^{-3/10})$ then the ABC posterior mean will have the same asymptotic distribution as for the case we consider, where $\epsilon_n=O(1/\sqrt{n})$. 
 
The results we stated  for the ABC posterior in section \ref{S:Posterior_sub} for the case $\epsilon_n=O(1/\sqrt{n})$ included
expressions for the posterior mean; see (\ref{eq:Mean1}) and (\ref{eq:Mean2}). The latter expression was under the assumption of a Gaussian kernel in ABC, but most of the exposition we give below holds for a general kernel \cite[see][for more details]{li2015}. 

The first of these, (\ref{eq:Mean1}), is the
true posterior mean given the summaries. Asymptotically our re-scaled observation $\tsobs$ has a Gaussian
distribution with mean 0 and variance $A(\btheta_0)$ due to the central limit theorem assumption, and the posterior
mean for $\bt$ is a linear transformation of $\tsobs$. This immediately gives
that the asymptotic distribution of the ABC posterior mean of $\bt$ is Gaussian with mean 0 and variance $I^{-1}$. Equivalently, for large $n$, the ABC posterior mean for $\btheta$ will be approximately normally distributed with mean $\btheta_0$ and variance $I^{-1}/n$. 
\index{ABC Posterior Mean Asymptotics}

The case where $\sqrt{n}\epsilon_n\rightarrow c$ for some $c>0$ is more interesting. If we have $d=p$, so we have the same number of summaries as we have parameters, then $D_0$ is a square matrix. Assuming this matrix is invertible, we see that
the ABC posterior mean simplifies to $D_0^{-1}\tsobs$. Alternatively if $d>p$ but $\Sigma=\gamma A(\btheta_0)$ for some scalar $\gamma>0$, so that the variance of our ABC kernel is proportional to the asymptotic variance of the noise in our summary statistics, then the ABC posterior mean again simplifies; this time to
\[
\left(D_0^TA(\theta_0)^{-1}D_0\right)^{-1}D_0^TA(\theta_0)^{-1}\tsobs.
\]
In both cases the expressions for the ABC posterior mean are the same as for the $c=0$ case, and are identical to the true posterior mean given the summaries. Thus the ABC posterior mean has the same limiting Gaussian distribution as the
true posterior mean in these cases.

More generally for the $c>0$ case, the ABC posterior mean will be different from the true posterior mean given the summaries. In particular the asymptotic variance of the ABC posterior mean can be greater than the asymptotic variance of the true posterior mean given the summaries. \cite{li2015} show that it is always possible to project a $d>p$ dimensional summary to a $p$ dimensional summary such that the asymptotic variance of the true posterior mean is not changed. This suggests using such a $p$ dimensional summary statistic for ABC \cite[see][for a different argument for choosing $d=p$]{fearnhead2012constructing}. An alternative conclusion from these results is to scale the distance used when deciding whether to accept or 
reject summaries to be proportional an estimate of the variance of the noise in the summaries.

It is interesting to compare the asymptotic variance of the ABC posterior mean to the limiting value of the ABC posterior variance. Ideally these would be the same, as that implies that the ABC posterior is correctly quantifying uncertainty. We do get 
equality when $\epsilon_n=o(1/\sqrt{n})$; but in other cases we can see that the ABC posterior variance is larger than
the asymptotic variance of the ABC posterior mean, and thus ABC over-estimates uncertainty. We will return to this in Section \ref{S:RegAdj}.
 \index{ABC Posterior Mean Asymptotics}

\section{Monte Carlo Error} \label{S:MonteCarlo}

The previous section included results on the asymptotic variance of the ABC posterior mean -- which gives a measure of accuracy of using the ABC posterior mean as a point estimate for the parameter. In practice we cannot calculate the ABC posterior mean analytically and we need to use output from a Monte Carlo algorithm, such as the rejection sampler described in the introduction. A natural question is what effect does the resulting Monte Carlo error have? And can we implement ABC in such a way that, for a fixed Monte Carlo sample size, the Monte Carlo estimate of the ABC posterior mean is an accurate point estimate? Or do we necessarily require the Monte Carlo sample size to increase as $n$ increases.

\cite{li2015} explore these questions. To do so they consider an importance sampling version \index{ABC Importance Sampler} of the rejection sampling algorithm we previously introduced. This algorithm requires the specification of a proposal distribution for the parameter, $q(\btheta)$, and involves iterating the following $N$ times 
\begin{itemize}
\item[(IS1)] Simulate a parameter from the proposal distribution: $\btheta_i \sim q(\btheta)$.
\item[(IS2)] Simulate a summary statistic from the model given $\btheta_i$: $\bs_i \sim f_n(\bs|\btheta_i)$.
\item[(IS3)] If $\lVert \sobs-\bs_i \rVert<\epsilon_n$ accept $\btheta_i$ and assign it a weight proportional to $\pi(\btheta_i)/q(\btheta_i)$.
\end{itemize}
The output is a set of, $N_{acc}$ say, weighted parameter values which can be used to estimate, for example, posterior means. With a slight
abuse of notation, if the accepted parameter values are denoted $\btheta^k$ and their weights $w_k$ for $k=1,\ldots,N_{acc}$ then we would estimate the posterior mean of $\btheta$ by
\[
\hat{\btheta}_N =\frac{1}{\sum_{k=1}^{N_{acc}} w_k} \sum_{k=1}^{N_{acc}} w_k\btheta^k.
\]
The use of this Monte Carlo estimator will inflate the error in our point estimate of the parameter by $\mbox{Var}(\hat{\btheta}_N)$, where we calculate variance with respect to randomness of the Monte Carlo algorithm.

If the asymptotic variance of the ABC posterior mean is $O(1/n)$ we would want the Monte Carlo variance to be $O(1/(nN))$. This would mean that the overall impact of the Monte Carlo error is to inflate the mean square error of our estimator of the parameter by a factor $1+O(1/N)$ \cite[similar to other likelihood free methods; e.g.][]{gourieroux1993indirect,Frigessi:2004}.

Now the best we can hope for with a rejection or importance sampler would be equally weighted, independent samples from the ABC posterior. The Monte Carlo variance of such an algorithm would be proportional to the ABC posterior variance. Thus if we want the Monte Carlo variance to be $O(1/n)$ then we need $\epsilon_n=O(1/\sqrt{n})$, as for slower rates the ABC posterior variance will decay more slowly than $O(1/n)$. 

Thus we will focus on $\epsilon_n=O(1/\sqrt{n})$. The key limiting factor in terms of the Monte Carlo error of our rejection or importance sampler is the acceptance probability. To have a Monte Carlo variance that is $O(1/n)$ we will need an implementation whereby the acceptance probability is bounded away from 0 as $n$ increases. To see whether and how this is possible we can examine the acceptance criteria in step, (RS3) or (IS3):
\[
\lVert \sobs-\bs_i \rVert = \lVert \{\sobs-\bb(\btheta_0)\} + \{\bb(\btheta_0)-\bb(\btheta_i)\} + \{\bb(\btheta_i)-\bs_i\}\rVert.
\]
We need this distance to have a non-negligible probability of being less than $\epsilon_n$. Now the first and third bracketed terms on the right-hand side will be $O_p(1/\sqrt{n})$ under our assumption for the central limit theorem for the summaries. Thus this distance is at best $O_p(1/\sqrt{n})$, and if $\epsilon_n=o(1/\sqrt{n})$ the probability of the distance
being less than $\epsilon_n$ should tend to 0 as $n$ increases. 

This suggests we need  $\sqrt{n}\epsilon_n\rightarrow c$ for some $c>0$. For this choice, if we have a proposal which 
has a reasonable probability of simulating $\btheta$ values within $O(1/\sqrt{n})$ of $\btheta_0$, then we could expect the
distance to have a non-zero probability of being less than $\epsilon_n$ as $n$ increases. This rules out the rejection sampler,
or any importance sampler with a pre-chosen proposal distribution. But an adaptive importance sampler that learns a good proposal distribution \cite[e.g.][]{sisson2007sequential,beaumont2009adaptive,peters2012sequential} can have this property.

Note that such an importance sampler would need a proposal distribution for which the importance sampling weights are also well-behaved. \cite{li2015} give a family a proposal distributions that have both an acceptance probability that is non-zero as $n\rightarrow\infty$ and have well-behaved importance sampling weights. 

Whilst \cite{li2015} did not consider MCMC based implementations of ABC \cite[]{Marjoram:2003,bortot2007inference}, the intuition behind the results for the importance sampler suggest that we can implement such algorithms in a way that the Monte Carlo variance will be $O(1/(nN))$. For example if we use a random walk proposal distribution with a variance that is $O(1/n)$ then after convergence the proposed $\btheta$ values will be a distance $O_p(1/\sqrt{n})$ away from $\btheta_0$ as required. Thus the acceptance probability should be bounded away from 0 as $n$ increases. Furthermore such a scaling is appropriate for a random walk proposal to efficiently explore a target whose variance is $O(1/n)$ \cite[]{roberts2001optimal}. Note that care would be needed whilst the MCMC algorithm is converging to stationarity as the proposed parameter values at this stage will be far away from $\btheta_0$.

\section{The Benefits of Regression Adjustment} \label{S:RegAdj} \index{Regression Adjustment}

We finish this chapter by briefly reviewing asymptotic results for a popular version of ABC which post-processes the output of ABC using regression adjustment. This idea was first proposed by \cite{Beaumont:2002} \cite[see][for links to Bayes linear methods]{nott2014approximate}. We will start with a brief description, then show how using regression adjustment can enable the adjusted ABC posterior to have the same asymptotic properties as the true posterior given the summaries, even if $\epsilon_n$ decays slightly slower than $1/\sqrt{n}$.

\begin{figure}
\centering
\includegraphics[scale=0.45]{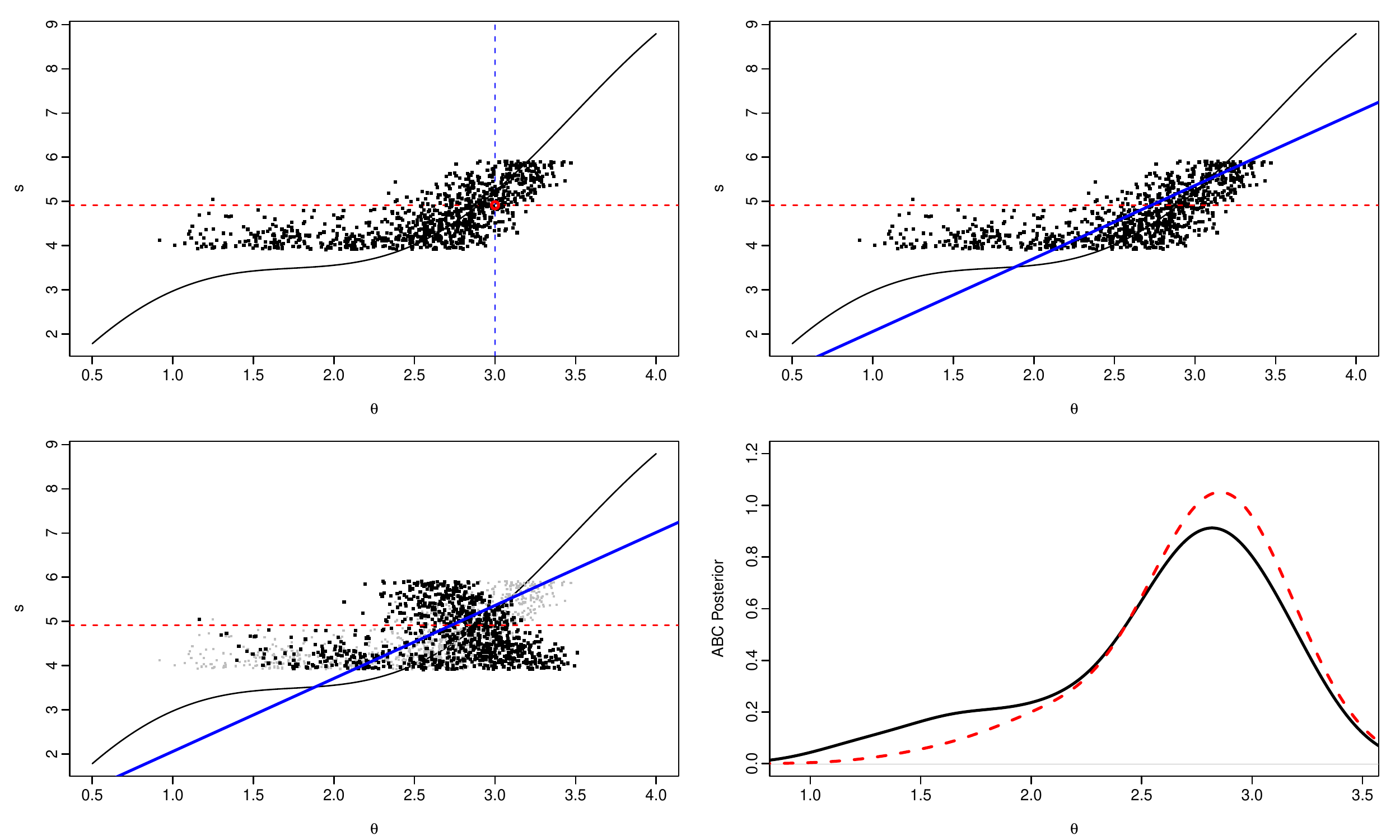}
\caption{
Example of the regression correction procedure of \cite{Beaumont:2002} for a single parameter, single summary statistic. Output of an ABC algorithm (top-left) showing accepted pairs of parameter and summary values (dots), the binding function for this model (solid black line), and $\theta_0$ and $\sobss$ (red circle and also blue vertical and red horizonal lines respectively). Top-right: the fit from a linear model predicting the parameter value from the summary (blue solid line). Bottom-left: the adjusted output (black dots; with original output in grey); we plot both old and adjusted parameter values against original summary statistic values.  Bottom-right: the ABC posterior based on the original accepted parameter values (black solid line) and the adjusted values (red dashed line).
\label{Fig:RegABC}
}
\end{figure}

Figure \ref{Fig:RegABC} provides an example of the ABC adjustment. The idea is to run an ABC algorithm that accepts pairs of parameters and summaries. Denote these by $(\btheta^k,\bs^k)$ for $k=1\ldots,N_{acc}$. These are shown in the top-left plot of
Figure \ref{Fig:RegABC}. We then fit $p$ linear models that, in turn, aim to predict each component of  the parameter vector from the summaries. The output of this fitting procedure  is  a $p$-dimensional vector $\hat{\balpha}$, the intercepts in the $p$ linear models, and a $p\times d$ matrix $\hat{B}$, whose $ij$th entry is the coefficient of the $j$ summary statistic in the linear model for estimating the $i$th component of $\btheta$. 

An example of such fit is shown in the top-left hand plot of Figure \ref{Fig:RegABC}. This fit is indicative of  biases in our accepted  $\btheta$ which correspond to different values of the summaries. In our example, the fit suggests that $\btheta$ values accepted for smaller, or larger, values of the summary statistic will, on average, be less then, or greater than, the
true parameter value. We can then use the fit to correct for this bias. In particular we can adjust each of the accepted parameter values, to $\tilde{\btheta}^k$ for $k=1,\ldots,N_{acc}$ where
\[
\tilde{\btheta}^k=\btheta^k-\hat{B}(\bs^k-\sobs).
\]
The adjusted parameter values are shown in the bottom-left plot of Figure \ref{Fig:RegABC}, and a comparison of the ABC posteriors before and after adjustment are shown in the bottom-right plot. From the latter we see the adjusted ABC posterior has a smaller variance and has more posterior mass close to the true parameter value.

The vector $\hat{\balpha}$ and the matrix $\hat{B}$ can be viewed as estimates of the vector $\balpha$ and the matrix $B$ that minimises the expectation of 
\[
\sum_{i=1}^p \left(\btheta_i-\balpha_i-\sum_{j=1}^d B_{ij} \pmb{S}_j\right)^2
\]
where expectation is with respect to parameter, summary statistic pairs drawn from our ABC algorithm.
\cite{li2016} show that if we adjust our ABC output using this optimal $B$ then, for any $\epsilon_n=o(n^{-3/10})$, the adjusted ABC posterior has the same asymptotic limit as the true posterior given the summaries. Obviously the asymptotic distribution of the mean of this adjusted posterior will also have the same asymptotic distribution as the mean of the true posterior given the summaries.

The intuition behind this result is that, asymptotically, if we choose $\epsilon_n=o(n^{-3/10})$, then our accepted samples will concentrate around the true parameter value. As we focus on an increasingly small ball
around the true parameter value, the binding function will be well approximated by the linear regression model we are fitting. Thus the regression correction step is able to correct for the biases
we obtain from accepting summaries that are slightly different from the observed summary statistics. From this intuition we see that a key requirement of our model, implicit within the assumptions needed for the 
theoretical result, is that the binding function is differentiable at the true parameter value: as such a differentiability condition is needed for the linear regression model to be accurate.

In practice we use an estimate $\hat{B}$, and this will inflate the asymptotic variance of the adjusted posterior mean by a factor that is $1+O(1/N_{acc})$, a similar effect to that of using Monte Carlo draws to estimate the mean. Importantly we get these strong asymptotic results even when $\epsilon_n$ decays more slowly than $1/\sqrt{n}$. For such a choice, for example $\epsilon_n=O(n^{-1/3})$, and with a good importance sampling or MCMC implementation, the asymptotic acceptance rate  
of the algorithm will tend to 1 as $n$ increases.

\section{Discussion}

The theoretical results we have reviewed are positive for ABC. If initially we ignore using regression adjustment, then the results suggest that ABC with $\epsilon_n=O(1/\sqrt{n})$ and with an efficient adaptive importance sampling or MCMC algorithm will have performance that is close to that of using the true posterior given the summaries. Ignoring Monte Carlo error, the accuracy of using the ABC posterior mean will be the same as that of using the true posterior mean if either we have the same number of summaries as parameters, or we choose an appropriate Mahalanobis distance for measuring the discrepancy in summary statistics. However, for this scenario the ABC posterior will over-estimate the uncertainty in our point estimate. The impact of Monte Carlo error will only be to inflate the asymptotic variance of our estimator by a factor $1+O(1/N)$, where $N$ is the Monte Carlo sample size. 

We suggest that this scaling of the bandwidth, $\epsilon_n=O(1/\sqrt{n})$,  is optimal if we do not use regression adjustment. Choosing either a faster or slower rate will result in Monte Carlo error that will dominate. One way of achieving this scaling is by using an adaptive importance sampling algorithm and fixing the proportion of samples to accept. Thus the theory supports
the common practice of choosing the bandwidth indirectly in this manner.

Also based on these results, we suggest choosing the number of summary statistics to be close to, or equal to, the number of parameters, and choosing a distance for measuring the discrepancy in summary statistics that is based on the variance of the summary statistics. In situations where there are many potentially informative summary statistics then one of the many dimension reduction approaches, that try to
construct low dimensional summaries that are information about the parameters, should be used \cite[e.g.][]{Wegmann:2009,fearnhead2012constructing,blum2013comparative,prangle2014semi}. \index{Summary Statistic Choice}

The results for ABC with regression adjustment are stronger still. These show that the ABC posterior and its mean can have the same asymptotics as the true ABC posterior and mean given the summaries. Furthermore this is possible with $\epsilon_n$ decreasing more slowly than $1/\sqrt{n}$, in which case the acceptance rate of a good ABC algorithm will increase as $n$ increases. These strong results suggest that regression adjustment should be routinely applied. One word of caution is that the regression adjustment involves fitting a number of linear-models to predict the parameters from the summaries. If a large number of summaries are used then the errors in fitting these models can be large \cite[]{fearnhead2012constructing} and lead to under-estimation of uncertainty in the adjusted posterior \cite[]{marin2016abc}. This again suggests using a small number of summary statistics, close or equal to the number of parameters.

Whilst the choice of bandwidth is crucial to the performance of ABC, and the choice of distance can also have an important impact on the asymptotic accuracy, the actual choice of kernel asymptotically has little impact. It affects the form of the ABC posterior, but does not affect the asymptotic variance of the ABC posterior mean (at least under relatively mild conditions).  

These asymptotic results ignore any ``higher-order" effects of the kernel that become negligible as $n$ gets large; so there may be some small advantages of one kernel over another for finite $n$, but these are hard to quantify. 
Intuitively the uniform kernel seems the most sensible choice -- as for a fixed acceptance proportion it accepts the summaries closest to the observed. 
Furthermore in situations where there is model error it is natural to conjecture that a kernel with bounded support, such as the uniform kernel, will be optimal. For such a case we want to only accept summaries that are $d_0+O(1/\sqrt{n})$, for some constant distance $d_0>0$, away from the observed summary (see Figure \ref{Fig:ModelError}). This is only possible for a kernel with bounded support. 
\index{Model Error}

{\bf Acknowledgements} This work was supported by EPSRC through the i-like programme grant. It also benefitted from discussions during the BIRS workshop on Validating and Expanding ABC Methods in February 2017.

\bibliographystyle{royal}
\bibliography{ABCasymptotics}

\end{document}